\documentclass{iaus}
\usepackage{graphicx}

\title[Magellanic Clouds pulsating variables] 
{Pulsating variable stars in the Magellanic Clouds}

\author[Gisella Clementini]   
{Gisella Clementini$^1$}

\affiliation{$^1$INAF Osservatorio Astronomico di Bologna, \\ Via Ranzani n. 1,
40127 Bologna, Italy\\ email: {\tt gisella.clementini@oabo.inaf.it} 
}

\pubyear{2008}
\volume{256}  
\pagerange{1--12}
\jname{The Magellanic System: Stars, Gas, and Galaxies}
\editors{Jacco Th. van Loon \& Joana M. Oliveira, eds.}
\begin{document}

\maketitle

\begin{abstract}
Pulsating variable stars can be powerful tools to study 
the structure, formation and evolution of galaxies. 
I discuss the role that the Magellanic Clouds' pulsating 
variables play in our understanding of the whole Magellanic System, in light of 
results 
on pulsating variables produced by extensive observing campaigns like 
the MACHO and OGLE microlensing surveys. 
In this context, I also briefly 
outline the promise of new surveys 
and astrometric missions 
which will target the Clouds in the near future.

\end{abstract}

\firstsection 
\section{Introduction}

The Large Magellanic Cloud (LMC) and the Small Magellanic Cloud (SMC) represent the nearest templates where we can study the stellar 
populations and galaxy interactions in detail, and they are where we set up and verify the 
astronomical distance scale.  
The pulsating variable stars can play a fundamental role in this context, offering
several advantages with respect to normal stars. The light variation caused by the periodic 
expansion/contraction of the surface layers makes the pulsating stars easier to recognize than normal 
stars, even when stellar crowding is severe. Their main parameter, the pulsation period, 
is measured with great precision, is unaffected 
by distance and reddening, and is directly related to intrinsic stellar quantities such 
as the 
star mass, radius, and luminosity.
Among pulsating variables, the Classical Cepheids (CCs) are the brightest stellar standard candles after Supernovae. The 
Period-Luminosity relation ($PL$), for which we celebrate this year the 100th anniversary of discovery 
by Henrietta Leavitt, 
makes them primary distance indicators in establishing the cosmic distance scale.
On the other hand, since pulsating variables of different types are in different 
evolutionary phases, they can be used to identify stellar components of different ages in 
the host system: the RR Lyrae stars and the Population II Cepheids (T2Cs) tracing the oldest 
stars ($t >$ 10 Gyrs); the Anomalous Cepheids (ACs) tracing the 
intermediate-age component ($\sim 4-8$ Gyrs); and the CCs tracing the
young stellar populations ($50-200$ Myrs).
The role of pulsating stars becomes 
increasingly  important in stellar systems like the Magellanic Clouds (MCs) where stars of different age and 
metal abundance share the same region of the color magnitude diagram (CMD).
The RR Lyrae stars, in particular, belonging to the oldest generation of stars, eyewitnessed 
the first epochs of their galaxy's formation and thus can provide hints on the early formation and 
assembling of the MCs system.

Our knowledge of the pulsating variable stars and the census of the MCs variables 
have made dramatic steps forward thanks to the microlensing surveys which, as a by-product, 
revealed and measured magnitudes and periods 
for thousands of variables in both Clouds. The overwhelming amount of information which these 
surveys have produced, not fully exploited yet, allowed for the first time to study 
the properties of primary distance indicators such as the CCs and the RR Lyrae stars 
based on statistically significant
numbers, as well as to reveal unknown features and new types of variables.
These topics are the subject of the present review.

\section{Position on the HR diagram and main properties}
\begin{figure}[b]
\begin{center}
 \includegraphics[width=4.4in]{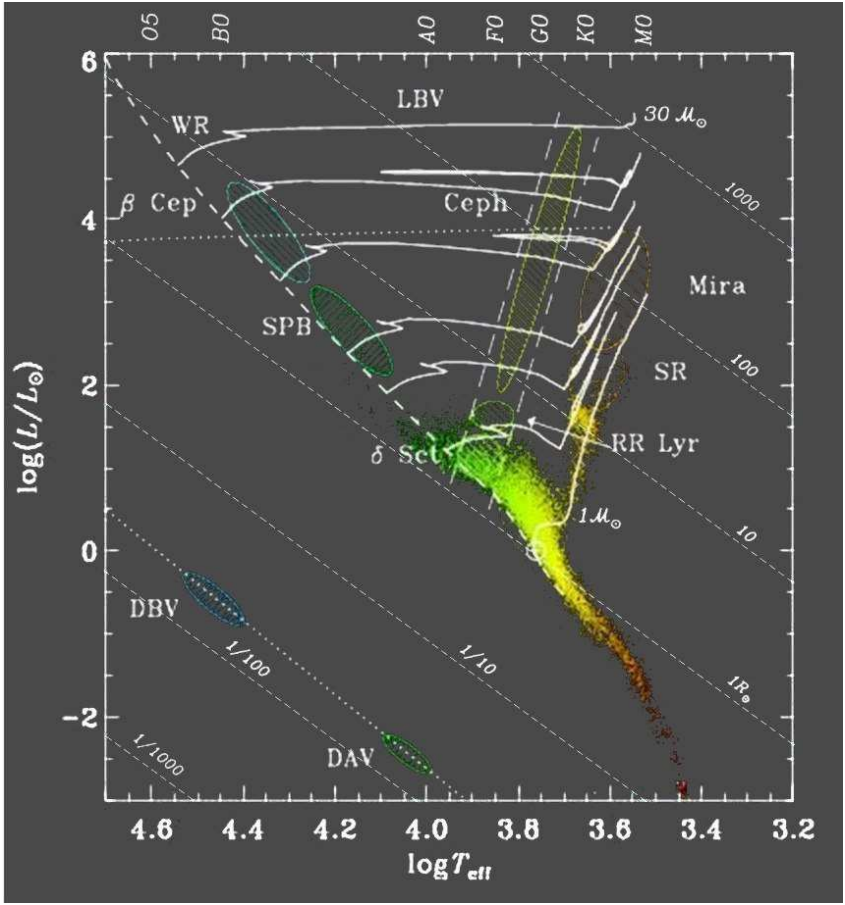} 
 \caption{Position of different types of pulsating stars in the HR diagram. The heavy dashed line stretching 
 from the upper left to the lower right is the main sequence of stars with solar abundances. 
 Lines of constant radius (from 1/1000 to 1000 $R_\odot$) 
  are shown, as well as 
 tracks corresponding to masses in the range from 1 to 30 $M_\odot$. The two long-dashed 
 lines 
 indicate 
 the position of the
 classical instability strip. Different acronyms mean: WR: Wolf-Rayet stars; LBV: luminous blue variables; 
 SPB: slowly pulsating B stars; SR: semiregular variables; DVB, DAV: pulsating 
 He white dwarfs (DB), and hydrogen-rich pulsating
 white dwarfs (DA).
 Adapted from \cite[Gautschy \& Saio (1995)]{GautschySaio1995}.}
   \label{fig1}
\end{center}
\end{figure}

Table\,\ref{tab1} presents an overview of the currently known major types of pulsating variables,
along with their main characteristics (typical period, absolute visual magnitude, 
parent stellar population and evolutionary phase). 
Fig.\,\ref{fig1} shows the loci occupied by different types of pulsating variables in the HR diagram, along with lines of constant
radius and a comparison with stellar evolutionary tracks for masses in the range from 1 to 30 M$_{\odot}$. 
The two long-dashed lines running almost vertically through
 the diagram mark the boundaries of  the so-called ``classical instability strip". 
Going from
 low to high luminosities in its domain we find:  $\delta$ Scuti variables 
 (luminosity: $log L/L_\odot \sim 1$, Spectral Type: A0-F0, mass: 
 M$\sim$ 2 M$_\odot$), the RR Lyrae stars ($log L/L_\odot \sim 1.7$, Spectral Type: A2-F2, 
 M$<$ 1 M$_\odot$), and the Cepheids (ACs: $log L/L_\odot \sim 2$, Spectral Type: F2-G6, 
 M$\sim$ 1-2 M$_\odot$; T2Cs: $log L/L_\odot \sim 2$, Spectral Type: F2-G6, 
 M$\sim$ 0.5 M$_\odot$; and CCs: $log L/L_\odot \sim 3-5$, Spectral Type: F6-K2, 
 M$\sim$ 3-13 M$_\odot$). Once extrapolated beyond the main sequence the instability strip crosses the region of the 
 pulsating hydrogen-rich DA white dwarfs. 
 Red variables (Miras and SRs) are situated instead below the red edge of the instability strip,
 at temperatures corresponding to spectral types K-M and luminosities $log L/L_\odot \sim 2-4$.
\begin{table}
  \begin{center}
  \caption{Main properties of different types of pulsating variable stars$^{\rm (1)}$}
  \label{tab1}
 {\scriptsize
  \begin{tabular}{|l|c|c|c|c|}\hline 
{\bf Class}                 & {\bf Pulsation Period}  & {\bf ${\rm M_V}$} & {\bf Population} & {\bf Evolutionary Phase} \\ 
                             &     (days)     &	 (mag)  	&		   &			        \\
\hline
$\delta$ Cephei (CCs)$^{\rm (2)}$         & 1$\div$100$^{(3)}$ & $-7 \div -2$      & I		  & Blue Loop		        \\
\hline
$\delta$ Scuti stars ($\delta$ Sc) & $<$0.5         & $2 \div 3  $      & I	  & MS - PMS 	                        \\
\hline
$\beta$ Cephei               & $<$0.3         & $-4.5 \div -3.5 $ & I		  & MS	                        \\
\hline
RV Tauri                     & 30$\div$100  & $-2 \div -1     $ & I,II		  & post-AGB 	                \\
\hline
Miras$^{\rm (4)}$                         & $100 \div 1000$       & $-2 \div  1     $ & I,II  	  & AGB 			\\
\hline
Semiregulars (SRs)$^{\rm (4)}$              & $>50$        & $-3 \div  1     $ & I,II  	  & AGB 			\\
\hline
RR Lyrae (RRL)               & 0.3$\div$1   & $ 0.0 \div 1    $ & II		  & HB	              	        \\
\hline
W Virginis (T2Cs)$^{\rm (5)}$   & 10$\div$50   & $-3 \div  1$ & II	          & post-HB		        \\
\hline
BL Herculis (T2Cs)$^{\rm (5)}$  & $<10$        & $-1 \div  0$ & II		  & post-HB		        \\
\hline
SX Phoenicis (SX Phe)        & $<0.1$       & $ 2 \div 3      $ & II		  & MS	                        \\
\hline
Anomalous Cepheids (ACs)      & $0.3\div 2.5$& $ -2 \div 0     $ & ?		  & HB turnover  	        \\
\hline
\end{tabular}
  }
 \end{center}
\vspace{1mm}
 \scriptsize{
 {\it Notes:}\\
  $^{(1)}$Adapted from (\cite[Marconi 2001]{Marconi_01}).\\
  $^{(2)}\delta$ Cephei variables are more commonly known as Classical Cepheids (CCs).\\
  $^{(3)}$ A few CCs with periods longer than 100 days are  known in both Clouds, in NGC~6822, NGC~55, NGC~300 and in IZw~18
  (\cite[Bird et al. 2008]{Bird_etal08}, and reference therein). Unfortunately, CCs with P$>$ 100 days are generally 
  saturated in the OGLE and MACHO photometry. They are now being observed with smaller telescopes in order to extend the $PL$ relation
   to longer periods (W. Gieren, private communication).\\
  $^{(4)}$ W Virginis and BL Herculis variables are often referred to as Population~II or Type~II Cepheids (T2Cs).\\
  $^{(5)}$ Miras and SRs often are jointly refereed to as red variables or long period variables (LPVs).}
\end{table}

\begin{figure}[b]
\begin{center}
 \includegraphics[width=5.0in]{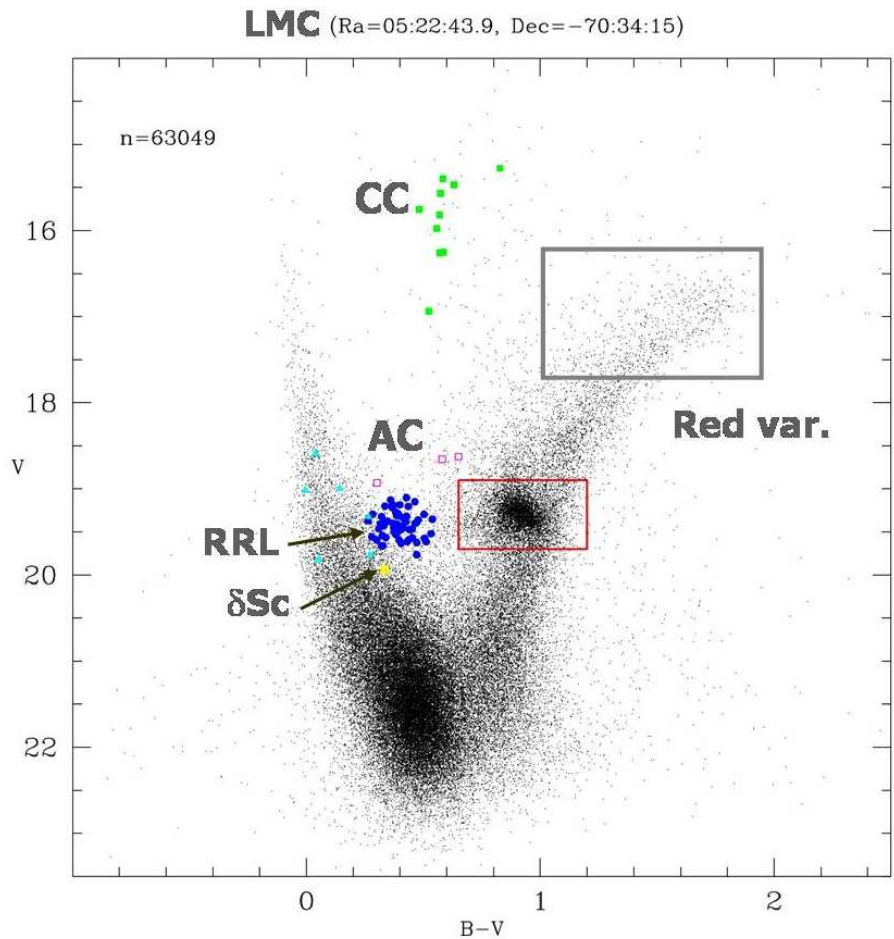} 
 \caption{Position of major types of pulsating stars in the CMD 
 of a region close to the LMC bar. Adapted from \cite{Clementini_etal03}.}
   \label{fig2}
\end{center}
\end{figure}
Given their complex stellar populations the MCs host samples of all the various types 
of pulsating stars shown in Fig.\,\ref{fig1}, although in varying proportions. 
As a combination of evolutionary/stellar population effects, 
but also due to the magnitude limit and time resolution of the currently available variability
surveys, up to now RR Lyrae stars, Cepheids and red variables 
are by far the most frequent and best studied variables in the MCs.
This is shown in Fig.\,\ref{fig2}, which displays the 
color magnitude diagram of a region close to the LMC bar, from the variability study 
of \cite{Clementini_etal03},  
with the different types of pulsating variables (Cepheids, RR Lyrae and $\delta$ Scuti stars) marked by 
different symbols, and the locus of red pulsating variables  
outlined by a large grey box. 

\section{The MCs pulsating variables in numbers}
Until the last decade of the twentieth century our knowledge of the MCs variables relied mainly 
upon  
photographic and photoelectric data and small and inhomogeneous samples.      
The situation drastically changed at the beginning of the nineties when the MACHO (http://wwwmacho.mcmaster.ca/) and EROS 
(http://eros.in2p3.fr/) microlensing experiments followed 
by OGLE (http://ogle.astrouw.edu.pl/) in 1997 began to regularly monitor the MCs for microlensing events and led to the discovery of 
thousands of pulsating stars.
Since then,
an increasing number of photometric surveys spanning the whole wavelength spectrum 
have taken a census of the
MCs pulsating variables and allowed to study in detail their pulsation properties on 
the basis of multiband light curves 
with hundreds of phase points spanning several years observations. 
Some of these surveys are summarized in Table\,\ref{tab2}. Also listed in the table
are the relatively few spectroscopic
studies available so far for the MCs pulsating variables and, in the last part of the table, 
the new photometric, spectroscopic and astrometric 
surveys which are planned for the next decade. 

In recent years the photometry of pulsating variables 
was progressively
 extended to the infrared 
region of the wavelength spectrum because  
the light variation of pulsating stars is smoother, reddening effects become negligible, and calibrating relations such 
as the $PL$ relations  become tighter when moving to the red region of the spectrum (see  
Figs.~4 and 6 of 
Madore \& Freedman 1991; Fouque {\it et al.} 2003; Ngeow \& Kanbur 2008; Freedman {\it et al.} 2008).
 The MCs variables make no exception, and 
several of the past and future
surveys listed in Table\,\ref{tab2} (e.g. ISO, DENIS, VMC) cover in fact the infrared spectral range
extending to the mid-infrared domain (3.6-8.0 $\mu$m) with the SAGE survey on the Spitzer satellite. 

The first results from the microlensing surveys of the MCs were published at the end of the nineties.
\cite{Alcock_etal96}, announced the discovery of about 
8000 RR Lyrae stars as a result of the MACHO microlensing survey of the LMC, and   
Alcock {\it et al.} (1998, 1999a, 1999b)  
reported on the identification of about 2000 LMC Cepheids. 
At same time, large numbers of Cepheids were found in the SMC by the EROS survey 
(Sasselov {\it et al.} 1997, Bauer {\it et al.} 1999), while 
\cite{Wood_etal99} discovered about 1400 variable stars (Miras, semi-regulars, contact and semi-detached binaries) 
defining five distinct parallel period-luminosity sequences 
on the red and asymptotic giant branches of the LMC. 
OGLE observations of the Clouds started only in 1997. However, the OGLE~II and III surveys represent the 
largest by area-coverage and the most deep and complete census of the MCs 
variables. First results from the OGLE~III survey of the LMC were recently published by \cite{Udalski_etal08}.
The new survey extends over about 40 square degrees and is about 
1-1.5 magnitudes deeper than the OGLE~II survey. 
Preliminary results on the LMC variables discovered by OGLE~III have been presented during this 
conference by I. Soszynski. 
They are summarized in Table \,\ref{tab3} and compared with results from the OGLE~II survey.
The number of LMC CCs has been almost doubled by OGLE~III, and significant numbers of ACs and $\delta$ Scuti stars
were also discovered (see Soszynski's talk in this Conference).  The increased sample of CCs traces very nicely the bar and 
gas-rich regions of the LMC. The RR Lyrae stars, instead, are evenly distributed on the whole field observed by 
OGLE~III and outline the more spheroidal distribution of the LMC's oldest stellar component.
The OGLE~III Wesenheit ($W_I$=$I-1.55(V-I)$) $PL$ diagram spans a total range of about 14 magnitudes,  
reaching about 1-1.5 magnitudes fainter 
than the OGLE~II $PL$ diagram (see Soszynski 2006). On this diagram the LMC $\delta$ Scuti stars locate on the extension to fainter magnitudes of the CCs $PL$, as 
suggested by \cite{McNamara_etal07}, based on the analysis of the very few $\delta$ Scuti stars which were known in the LMC.

In the following, I will specifically address some major results which the  surveys listed in Tables \,\ref{tab2} and  \,\ref{tab3} have 
produced in the study of Cepheids, RR Lyrae stars and red variables 
 and, in turn,
in our knowledge  of the 
MCs system.  
\begin{table}
  \begin{center}
  \caption{Surveys of MCs variable stars}
  \label{tab2}
 {\scriptsize
  \begin{tabular}{|l|c|c|c|}\hline 
{\bf Visual} & {\bf Infrared} & {\bf Spectroscopy} & {\bf Astrometry} \\ 
\hline
EROS$^{\rm (1)}$         & ISO                         &Luck \& Lambert (1992)        & \\
\hline
MACHO$^{\rm (2)}$        & DENIS                       &Luck {\it et al.} (1998)            & \\
\hline
SUPERMACHO$^{\rm (3)}$   & SIRIUS                      &Romaniello {\it et al.} (2005, 2008)& \\
\hline
OGLE II - III$^{\rm (4)}$& 2MASS                       &Gratton {\it et al.} (2004)         & \\
\hline
MOA$^{\rm (5)}$          & SAGE$^{\rm (6)}$            &Borissova {\it et al.} (2004, 2006) & \\
\hline
\hline
STEP@VST$^{\rm (7)}$     & VMC $^{\rm (8)} $            &STEP@VLT                       & \\ 
\hline
Gaia$^{\rm (9)}$         &                             &Gaia                           & Gaia\\
\hline
LSST$^{\rm (10)}$         &                             &                               & \\
\hline
\end{tabular}
  }
 \end{center}
\vspace{1mm}
\scriptsize{
{\it Notes:}\\
 $^{(1)}$ http://eros.in2p3.fr/\\
 $^{(2)}$ http://wwwmacho.mcmaster.ca/\\
 $^{(3)}$ http://www.ctio.noao.edu/\~\,supermacho/\\
 $^{(4)}$ http://ogle.astrouw.edu.pl/\~\, ogle/; http://bulge.astro.princeton.edu/\~\, ogle/\\ 
 $^{(5)}$ http://www.phys.canterbury.ac.nz/moa/\\
 $^{(6)}$ http://sage.stsci.edu/index.php\\
 $^{(7)}$ http://vstportal.oacn.inaf.it/node/40/\\ 
 $^{(8)}$ http://star.herts.ac.uk/\~\, mcioni/vmc/\\ 
 $^{(9)}$ http://gaia.esa.int\\
 $^{(10)}$ http://www.lsst.org/lsst\_home.shtml\\
}
\end{table}

\begin{table}
  \begin{center}
  \caption{Number of pulsating stars in the OGLE surveys of the MCs}
  \label{tab3}
 {\scriptsize
  \begin{tabular}{|l|c|c|c|c|}\hline 
{\bf Class}  	    & {\bf LMC } &  {\bf SMC } & {\bf Survey}& Reference \\ 
\hline
CCs          	    & 1416	 &2144         & OGLE~II     &  Udalski {\it et al.}(1999a,b,c)\\
                    &            &             &             &  Soszynski {\it et al.} (2000)  \\
             	    & 3361	 &	       & OGLE~III    &  Soszynski (this Conference)    \\
\hline
T2Cs       	    & 14	 &	       & OGLE~II     &  Kubiak \& Udalski (2003)       \\
\hline
ACs          	    & Yes	 &	       & OGLE~III    &  Soszynski (this Conference)    \\
\hline
RR Lyrae stars      & 7612	 &571	       & OGLE~II     &  Soszynski {\it et al.} (2002, 2003)\\
\hline
Miras \& SRs        & 3221       &             & OGLE~II     &  Soszynski {\it et al.} (2005)  \\
\hline
Small Amplitude Red Giants (SARGs)               & 15400       &3000            & OGLE~II - III    &  Soszynski {\it et al.} (2004, 2005)  \\
\hline
$\delta$ Scuti stars&            &             & OGLE~II     &           \\
\hline
                    & Yes        &             & OGLE~III    &  Soszynski (this Conference)     \\
\hline 
\end{tabular}
  }
 \end{center}
\end{table}

\subsection{The $PL$ relation of Classical Cepheids}
The CCs are primary standard candles that allow to link the local distance scale to the cosmological distances
needed to determine the Hubble constant, H$_0$.

The CCs $PL$ relation, discovered 
by H. Leavitt at the beginning of the twentieth century as she was picking up variables on photographic 
plates of the MCs, is unquestionably one of the most powerful tools at our disposal 
for determining the extragalactic distance scale. 
The extraordinary large number of CCs discovered in the MCs by the MACHO and OGLE surveys,
allowed to derive the $PL$ relations on unprecedented statistically significant and homogeneous samples of CCs.
\cite{Udalski_etal99a} used fundamental-mode (FU) CCs in the LMC and SMC to 
derive the following $PL$ relations:
$$V_0{\rm (LMC)}=-2.760 log P - 17.042, ~~~~ \sigma = 0.159 ~{\rm mag}, ~~~~ (649~~~{\rm FU~~~CCs})$$
$$V_0{\rm (SMC)}=-2.760 log P - 17.611, ~~~~ \sigma = 0.258 ~{\rm mag}, ~~~~ (466~~~{\rm FU~~~CCs})$$
The slope of these relations is in very good agreement with the slope of theoretical 
$PL$ relations computed by \cite{Caputo_etal00} from nonlinear convective pulsation models of CCs 
(${\rm M_V} = -2.75 log P - 1.37, ~~\sigma = 0.18$ mag).

In order to use the $PL$'s of the MCs CCs to 
measure distances outside the Clouds the zero point of the $PL$  relation is generally 
fixed by using Galactic CCs whose absolute magnitudes are known from parallax measurements and/or Baade-Wesselink analyses 
or, alternatively, by assuming a value for the distance to the LMC. In the latter case, the zero-point problem thus shifts to 
the problem of having a robust distance determination for the LMC.  
The HST key program (Freedman {\it et al.} 2001) used the slope of the CCs $PL$ relations by
\cite{Udalski_etal99a}  
and a zero-point consistent with an assumed true distance modulus for the LMC of $\mu_{\rm LMC}$=18.5 mag 
to measure distances to 31 galaxies with distances from 700 Kpc to 20 Mpc. These then served to calibrate other, 
more far-reaching secondary distance indicators to determine the Hubble constant in a region of constant Hubble flow
(see Freedman {\it et al.} 2001, but also Saha {\it et al.} 2001, and Tammann {\it et al.} 2008, for different
conclusions on the value of H$_0$).

In spite of the success in measuring distances up to 20 Mpc, a number of basic questions concerning the CCs $PL$ relation still need an answer (see Fouque {\it et al.} 2003, for a nice review on this topic). 
 Is the CCs $PL$ relation universal, as suggested, for instance, by \cite{Fouque_etal07}, so that we are allowed to apply the LMC $PL$ to CCs 
 in other galaxies?
 Does it depend on metal abundance, as also suggested by nonlinear pulsation models (see, e.g., Marconi {\it et al.} 2005, Bono {\it et al.} 2008, and
 references
 therein)? Is it linear or does it break at periods 
 around 10 days, as a number of studies (Tammann \& Reindl 2002, Tammann {\it et al.} 2002, 
 Kanbur \& Ngeow 2004, Sandage {\it et al.} 2004, Ngeow {\it et al.} 2005, 2008) are  now 
 suggesting?  And, how reliably do we know the distance to the LMC and the distance modulus of $\mu_{\rm LMC}$=18.5 mag 
adopted by the HST key program? I will try to address this latter question in Section 4.
\cite{Romaniello_etal08} provide a summary of the rather controversial results on the metallicity sensitivity 
of the Cepheid distances, in the literature of the past twenty years.    
These authors use direct spectroscopic measurements of iron abundance for 
Galactic and MC Cepheids to study the metallicity sensitivity of the CCs $PL$ and conclude that 
the $V$-band $PL$ is metallicity dependent, while no firm conclusions can be reached for the $K$-band $PL$. However,
in their recent paper based on 
OGLE~II and SAGE observations of the LMC CCs, \cite{Neilson_etal08}  find that the infrared $PL$ relations as well have additional 
uncertainty due to a metallicity dependence. 
  Clearly, elemental abundance estimates for larger numbers of CCs spanning broad metallicity ranges, increased samples of photometric data in
the infrared domain, and a fine tuning of all the parameters involved in the definition of the CCs $PL$ relations 
(see, e.g.  W. Gieren's talk, in this Conference) are needed to quantitatively assess  
the metallicity dependence of both zero-point and slope of the CCs $PL$ relations. 
Hopefully, most of the questions still pending on the CCs $PL$  relation will find more definite answers from the 
new surveys of the MCs variables planned for the next decade (see Section 5).

\subsection{The MCs  RR Lyrae stars}
The RR Lyrae stars are the primary distance indicators for stellar systems  mainly composed by an old stellar component.
They follow an absolute magnitude-metallicity relation in the visual band: ${\rm M_V} - {\rm [Fe/H]}$ (Sandage 1981a,b) and a tight ($\sigma \sim $ 0.05 mag) 
Period-Luminosity-Metallicity  relation in the $K$ band: $PL_{K}Z$, (Bono {\it et al.} 2003, Catelan {\it et al.} 2004, Sollima {\it et al.} 2008, and
references therein). Some observational and theoretical studies (see, e.g., Bono {\it et al.} 2003, Di Criscienzo {\it et al.} 2004, and references 
therein) have suggested that the ${\rm M_V} - {\rm [Fe/H]}$ relation is not linear, becoming steeper when moving to larger metal content.

RR Lyrae stars have been found in all Local Group (LG) galaxies irrespective
of morphological type and, although much fainter than the CCs, have been observed and measured as far as in the Andromeda galaxy. 
\cite{Alcock_etal96} discovered about 
8000 RR Lyrae stars in the LMC, among which a fairly large number of double-mode pulsators
(Alcock {\it et al.} 1997).  
Results from the 
study of the light curves (Alcock {\it et al.} 2000, 2003, 2004) 
showed that the average periods 
of the LMC RR Lyrae stars differ from what
is observed for the Milky Way (MW) variables. 
Alcock {\it et al.}'s results were later confirmed and strengthened by the OGLE~II studies of the LMC and SMC RR Lyrae stars 
(Soszynski {\it et al.} 2002, 2003). These findings suggest differences in the star 
formation history and rule out both MCs as possible contributors to the assembling of the MW halo.

Spectroscopic data of about 250 LMC RR Lyrae were obtained by \cite{Gratton_etal04}, and Borissova {\it et al.}
(2004, 2006). They were used to estimate the metal abundance, radial velocity and radial velocity dispersion of the LMC RR Lyrae population. 
The LMC RR Lyrae stars are metal-poor, with average metal abundance of
$\langle {\rm [Fe/H]} \rangle = -1.48/1.54$ dex and spread of about 0.2-0.3 dex. The 
radial velocity dispersion, $\sigma_{v_r}$=50 km/s, does not vary significantly with increasing distance from the LMC center 
(Borissova {\it et al.} 2006 and references therein), and is higher than the velocity dispersion of
any other LMC population previously measured, thus providing empirical evidence for a kinematically hot, metal-poor halo  in the LMC.
\cite{Gratton_etal04} combined their spectroscopic measurements with high accuracy $V$  magnitudes for about a hundred
RR Lyrae stars by \cite{Clementini_etal03} 
to derive the luminosity-metallicity relation of the LMC variables, for which 
they found the following linear relation: ${\rm M_V}$ = 0.214 ([Fe/H] + 1.5) + 19.064. The slope of this relation agrees very well 
with slopes derived for the luminosity-metallicity relation of the MW RR Lyrae stars (Fernley {\it et al.} 1998) 
and horizontal branch stars in the globular clusters of M31 (Rich {\it et al} 2005), thus supporting the idea that 
the luminosity-metallicity relation of the RR Lyrae stars is, in first approximation, linear and universal.

Both the ${\rm M_V} - {\rm [Fe/H]}$ and the $PL_{K}Z$ relations were extensively used to measure distances to the
LMC field and globular cluster's stars (see, e.g. Clementini {\it et al.} 2003, Dall'Ora  {\it et al.} 2004, Szewczyk 
{\it et al.} 2008).
Results from these studies are summarized in Table \,\ref{tab4}. 

\subsection{The Red Variables}

The red variables are, typically, highly evolved stars in the Asymptotic Giant Branch (AGB) phase. Their atmospheres pulsate with typical 
periods in the range from several ten to several hundreds days, and 
amplitudes ranging from 0.1 up to 6 magnitudes. The class 
 includes the first ever recorded pulsating star: Mira, the
prototype of variables with the largest visual amplitudes of any class of pulsating stars: the Miras.
The light curves of these variables
 are often semiregular and multiperiodic, with 
short brightness outbursts observed sometimes on top of the periodic light change. Mass loss and dust emission, typical of 
the AGB evolutionary phase, further complicate the scenario, and even the mode of radial pulsation of these stars has long remained a matter of
debate. Although the study of these variables most benefited from the long-term photometric monitoring of the Clouds by the microlensing surveys,  
 and then from the combination of the visual data with infrared photometry, still they remain perhaps the least understood of all variable stars.

\cite{Wood_etal99} found that about 1400 red and asymptotic giant branch stars observed in the LMC by the MACHO survey were 
long period variables. They identified 5 distinct parallel $I$-band $PL$ sequences, (labeled from ``A" to ``E" in their Fig. 1), 
 and derived 
a first tentative classification of the red variables. 
By combining the MACHO photometry to infrared $J$ and $K$ data \cite{Wood00} further refined this classification,  
and definitely identified the Miras as fundamental mode pulsators falling on a single $PL_K$ relation corresponding to sequence ``C" of  
\cite{Wood_etal99}. The SR variables are instead  first to third overtone, or even 
fundamental mode pulsators falling on sequence ``B", the small amplitude red variables are on sequence ``A", the
long secondary period variables on sequence ``D", and, finally, the contact binaries are on sequence ``E".
This classification was confirmed by various authors (e.g. Lebzelter {\it et al.} 2002). 
Since the Miras are bright, large amplitude variables, their $PL$ relation is an important distance indicator for old and 
intermediate age populations. A new calibration of the Miras $PL_K$ relation was recently derived by \cite{Whitelock_etal08} 
using 53 LMC Miras with periods less than 420 days.

 The number of red variables identified around the tip of the MCs red and asymptotic giant branches has 
 massively increased in the last years. \cite{Fraser_etal05} detected about 22000 red variables by combining the 8 year light-curve database from the MACHO survey 
 of the LMC, with  2MASS infrared $J,H,K$  photometry. The OGLE~II and III surveys detected more than 3000 Miras and SR 
 variables in the
LMC (Soszynski {\it et al.} 2005), and about 15400 and 3000 small amplitude red giant variables (SARGs) respectively in the LMC and in the SMC.
These variables appear to be a mixture of AGB and red giant branch
 pulsators (Soszynski {\it et al.} 2004).
Ita {\it et al.} (2004a,b) combining results from the OGLE~II and the SIRIUS near-infrared $JHK$ surveys, found that variable red giants in 
the SMC form parallel sequences  
on the $PL_K$ plane, just like those found by Wood in the LMC. Moreover, Wood's original sequences were found to split into several separate subsequences
above and below the tip of the LMC and SMC red giant branches. Slightly different relation were also found for carbon- and oxygen-rich variables. 
The number of $PL$ sequences identified in the red variable domain was brought to fourteen by \cite{Soszynski_etal07}, who also found that the 
slopes of the $PL$ relation for Miras and SR variables seem to be the same in the LMC and SMC. The number of $PL$s is expected to increase further
once the analysis of the red variables in the OGLE~III database will be completed.   

\section{The distance to the LMC}
Because the LMC is the first step of the extragalactic distance ladder, the knowledge of 
its distance has a tremendous impact on the entire astronomical distance scale. 
\cite{Benedict_etal02} published an historical summary 
of distances to the LMC from different indicators. Their Fig. 8  provides an impressive overview  of 
the dispersion in the LMC distance moduli ($\mu_{\rm LMC}$) 
published during the ten-year span from 1992 to 2001.
The last decade has 
seen dramatic progress in the calibration of the different distance indicators. The 
dispersion in $\mu_{\rm LMC}$ has definitely shrunk, and values at the extremes of Benedict {\it et al.}'s distribution 
(18.1 and 18.8 mag, respectively) 
are not seen very often in the recent literature. 

Table \ref{tab4} summarizes some recent determinations of LMC
distances based on pulsating variables.
Far from being exhaustive, this table is only meant to highlight some recent 
advances in the distance determinations based on major types of pulsating variables found in the LMC.
Although systematic differences still exist and need to be worked upon, (for instance 
the metallicity-corrected $PL$ relation based on revised Hipparcos parallaxes for CCs, van Leeuwen {\it et al.} 2007,
gives a somewhat shorter modulus, 
as does also 
the $PL$ based on new values for the $p$-factor used to transform radial velocities into
pulsational velocities in the Baade-Wesselink analyses of CCs, see Gieren's talk in this Conference),
the controversy between
the so-called ``short" and ``long" distances to the LMC seems to have largely vanished, and 
there is now a substantial 
convergence of the most reliable standard candles on a distance modulus for the LMC  
around 18.5 mag (Clementini {\it et al.} 2003, Walker 2003, Alves 2004, Romaniello {\it 
et al.} 2008).
\begin{table}
  \begin{center}
  \caption{Distances to the LMC from pulsating variables}
  \label{tab4}
 {\scriptsize
  \begin{tabular}{|l|c|c|}\hline 
{\bf Method} & {\bf Distance modulus} &  {\bf Reference } \\ 
\hline
$PL$, LMC $\delta$ Scuti stars         & 18.50 $\pm$ 0.22  &McNamara {\it et al.} (2007)\\
\hline
Model fitting, $\delta$ Scuti stars   & 18.48 $\pm$ 0.15  &McNamara {\it et al.} (2007)\\
\hline
Model fitting, Bump Cepheids          & 18.48 $\div$ 18.58       &Bono {\it et al.} (2002)     \\
                                     & 18.55  $\pm$ 0.02 &Keller \& Wood (2002)\\            
                                     & ~18.54  $\pm$ 0.018&Keller \& Wood (2006)\\            
\hline
Model fitting, field RR Lyrae stars         & 18.54  $\pm$ 0.02 &Marconi \& Clementini (2005)\\
\hline
${\rm M_V} - {\rm [Fe/H]}$, field RR Lyrae stars     & 18.46  $\pm$ 0.07$^{(1)}$ &from Clementini {\it et al.} (2003)\\
\hline
$PL_{\rm K}Z$, field RR Lyrae stars & 18.48$\pm 0.08$    &  Borissova {\it et al.} (2004)\\
\hline
$PL_{\rm K}Z$, RR Lyrae stars in Reticulum& 18.52$\pm 0.01 \pm$ 0.12&  Dall'Ora {\it et al.} (2004)\\ 
\hline
$PL_{\rm K}Z$, field RR Lyrae stars & 18.58$\pm 0.03 \pm$ 0.11&  Szewczyk {\it et al.} (2008)\\
\hline
$PL_K$, CCs                &   18.47$\pm 0.03$                      &  van Leeuwen {\it et al.} (2007)\\
$PLC_{J,K}$, CCs           &   18.45$\pm 0.04$                      &                                 \\
$PL_W$, CCs                &   18.52$\pm 0.03$                      &                                 \\
$PL_W$, CCs                &   18.39$\pm 0.03$                      &  metallicity-corrected     \\
\hline
\end{tabular}
  }
 \end{center}
\vspace{1mm}
\scriptsize{
{\it Notes:}\\
$^{(1)}$ This distance modulus was derived using values from \cite{Clementini_etal03}for  $\langle V(RR) \rangle$ and the reddening,
and the assumption of $M_V$=0.59 $\pm 0.03$ mag for the absolute visual magnitude of RR Lyrae stars at metal abundance 
[Fe/H]=$-1.5$ (Cacciari \& Clementini 2003).\\}
\end{table}

\section{The new surveys}
Among the new photometric surveys, STEP and VMC, expected to start at the end of 2009, will  
repeatedly observe the Clouds allowing to study variable stars.

STEP (The SMC in Time: Evolution of a Prototype interacting dwarf galaxy, 
see the poster contribution to this Conference by  Ripepi {\it et al.}), 
will use the VLT Survey Telescope
(VST) to obtain $V, B$ and $i^{\prime}$ single-epoch photometry 
of the SMC, reaching below the galaxy main sequence turn-off; 
as well as shallow 
time-series photometry of the Wing and Bridge toward the LMC, for which no previous 
variability survey exists yet, reaching variables as faint as the RR Lyrae stars.

VMC (VISTA near-infrared survey of the Magellanic Clouds, see the poster paper by  Cioni {\it et al.} in this Conference) is instead an ESO 
public survey, which will obtain near-infrared $YJK_s$ photometry of the whole Magellanic System (LMC, SMC and Bridge) 
with the subset of $K_s$ exposures taken in time-series fashion 
to study variable stars.   

These two surveys together will provide new multiband data to study the spatially resolved star formation history
of the MCs, and will allow to reconstruct the 3-dimensional structure of 
the whole Magellanic System using various types of pulsating stars 
(Classical, Type II and Anomalous Cepheids, RR Lyrae, $\delta$ Scuti, and Miras). 

The astrometric satellite Gaia, planned for launch in 2011, is one of the European Space Agency (ESA) cornerstone missions.
During its lifetime of nominally 5 years, Gaia will scan the entire 
sky repeatedly, with an average frequency of about 80 measurements per object over the five-year time span, and  
will provide astrometry, 2-color photometry,  and slit-less spectroscopy in the Ca triplet domain 
($847-874$ nm) for all
sources brighter than $V \sim $20 mag (about 1.3 $\times$ 10$^9$ stars in total).
Expected errors of the Gaia measurements are: $\sigma_{\pi}$=10-25 $\mu$arcsec at $V\sim$15 mag for parallaxes, and 
$\sigma$  = 15 km/s at $V <6-17$ mag for radial velocities. This is the domain of the bright pulsating variables in the MCs, which, 
if the satellite performs as expected, will then have their parallax, magnitude, radial velocity and metal
abundance directly measured by Gaia. The direct measure via trigonometric parallaxes of distances for Magellanic Cloud CCs and Miras
will thus allow the 
calibration with unprecedented precision of these most important primary standard candles.
\section{Acknowledgments}
It is a pleasure to thank Marcella Marconi for comments and suggestions on a preliminary version of this review, and Thomas Lebzelter for useful 
directions on the literature of the red variables. A special thanks goes to the editors, Jacco van Loon \& Joana Oliveira
for patiently waiting for my Conference Proceedings.\\

\end{document}